\documentclass[aps,prl,showpacs,twocolumn]{revtex4-1}

\usepackage{graphicx}

\begin{document}

\title{Nonequilibrium Phase Transition of Interacting Bosons \\ in an Intra-Cavity Optical Lattice}

\author{M. Reza Bakhtiari$^1$, A. Hemmerich$^2$, H. Ritsch$^3$, and M.
Thorwart$^1$}
\affiliation{$^1$I.\ Institut f\"ur Theoretische Physik, Universit\"at
Hamburg,
Jungiusstra{\ss}e 9, 20355 Hamburg, Germany \\
$^2$Institut f\"ur Laser-Physik, Universit\"at Hamburg, Luruper
Chaussee 149, 22761 Hamburg, Germany \\
$^3$Institute for Theoretical Physics, Universit\"at Innsbruck,
Technikerstra{\ss}e 25, 6020 Innsbruck, Austria }

\begin{abstract}

We investigate the nonlinear light-matter interaction of a Bose-Einstein condensate trapped in an external periodic potential inside an optical cavity, which is weakly coupled to the vacuum radiation modes and driven by a transverse pump field. Based on a generalized Bose-Hubbard model, which incorporates a single cavity mode, we include the collective back action of the atoms on the cavity light field and determine the nonequilibrium quantum phases within the non-perturbative bosonic dynamical mean-field theory. With the system parameters adapted to recent experiments, we find a quantum phase transition from a normal phase to a self-organized superfluid phase, which is related to the Hepp-Lieb-Dicke superradiance phase transition. For even stronger pumping, a self-organized Mott insulator phase arises.

\end{abstract}

\bibliographystyle{prsty}

\maketitle
High finesse optical cavities are nowadays routinely used in a wide
range of fundamental and applied research \cite{Vah:03} as a powerful tool 
to investigate the interaction of matter and light. Thus, the 
celebrated Jaynes Cummings (JC) model of a two-level 
atom coupled to a single mode of 
the radiation field \cite{Aga:12} turned into a precise description of an
experimental scenario that can be implemented in the laboratory
\cite{Har:06}. Even more, modern optical cavities provide a tool to
investigate \textit{collective} matter-light interaction beyond a
single-atom picture. An early insightful theoretical proposal
\cite{Dom:02} and a subsequent experiment \cite{Bla:03} have shown that in a
thermal atom ensemble in a high-finesse optical cavity, two
different phases can emerge depending on the strength of an external pump
laser field: a normal phase, characterized by a homogeneous density
distribution and a vanishing intra-cavity field, and a self-organized (SO)
phase, where the atoms form a matter grating trapped by a stationary
intra-cavity optical standing wave. 
Only recently, similar experiments were realized with Bose-Einstein
condensates (BECs) \cite{Bau:10, Kes:14, Kli:14}. It was realized that in
the limit of zero temperature and moderate external pumping this scenario
represents an experimental implementation of the open Dicke model
\cite{Dic:54}. The normal-to-SO phase transition is directly related
to the quantum phase transition early discovered in the Dicke model by
Hepp and Lieb \cite{Hep:73}. These results have triggered a tremendous
revisit of the Dicke model \cite{Rit:13}.   

\begin{figure}
\includegraphics[scale=0.40, angle=0, origin=c]{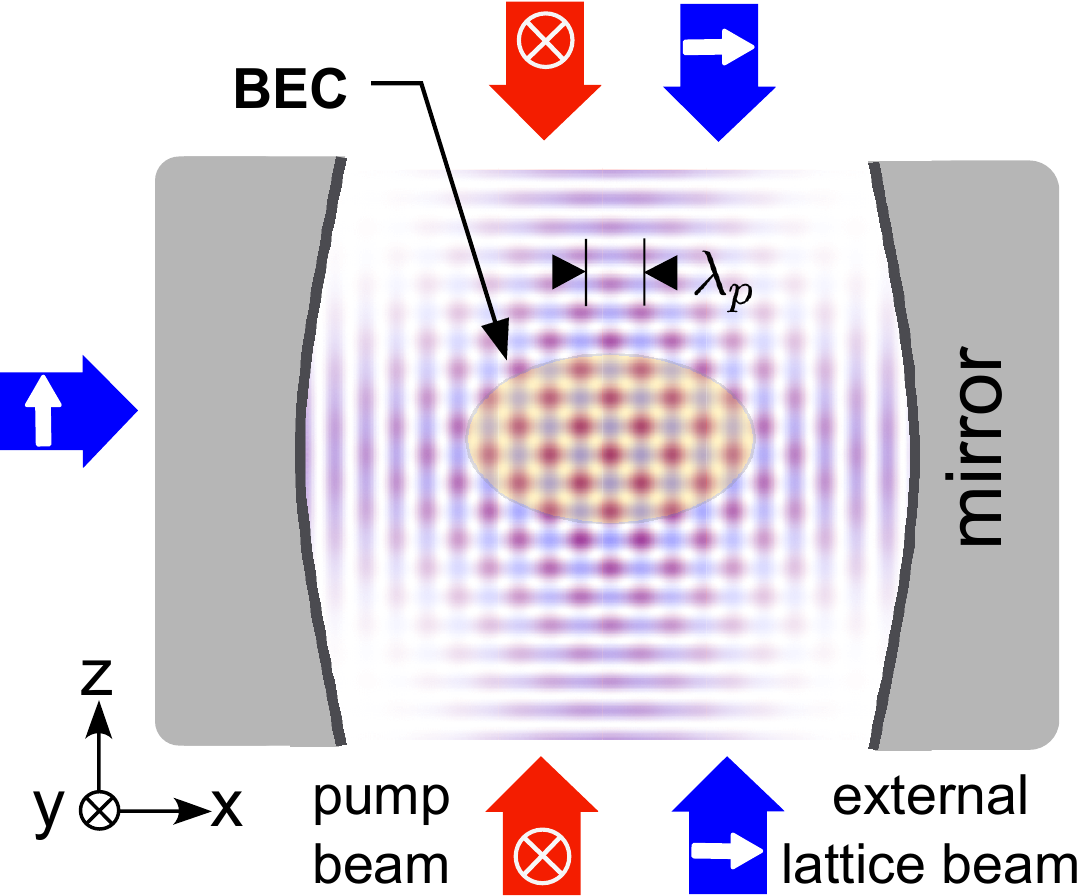}
\caption{Experimental setup. The external laser beams (blue) form a
square lattice in the $xz$-plane with a lattice constant $\lambda_{p}/2$ 
(blue pattern). The linear polarization of the pump beams (red) along the
$y$-direction allows scattering of photons into the cavity. This leads to
a square lattice in the $xz$-plane rotated with respect to the external
lattice by $45^{\circ}$ with the lattice constant $\lambda_{p}/\sqrt(2)$ (red
pattern). Each minimum of the pump-induced lattice potential coincides with a
minimum of the external lattice potential. The wavelength of all beams is
$\lambda_{p}$.}
\label{fig:Fig.1}
\end{figure}

An intriguing perspective is the extension of the present atom-cavity
experiments to study lattice models with cavity mediated infinite range
interactions. Experimentally this amounts to amending the BEC by an
external periodic potential giving rise to an optical lattice. Depending
on the choice of the spatial periodicity and phase of this additional
lattice potential, very different physical scenarios can arise. A most
natural choice is commensurability of the external lattice and the
pump-induced lattice formed by photon scattering, such that a potential
minimum of the pump-induced intra-cavity lattice potential corresponds to
a minimum of the external lattice potential. However, conversely, 
not necessarily every minimum of the external potential coincides with a minimum 
of the intra-cavity lattice potential. 
In this case the pump-induced lattice acts to deepen every second
minimum of the external lattice. This specific choice, studied in the
present work, is compatible with the present experiments, and can be
implemented by coupling laser beams of similar frequencies but different
polarizations to the cavity 
as illustrated in Fig.~\ref{fig:Fig.1}.

A comprehensive theory of the aforementioned physical
scenario requires to combine two paradigmatic models. The Bose-Hubbard
(BH) model has to be joined with the JC model 
to derive a generalized BH model \cite{Mas:05, Mas:08, Lar:08} (hereafter,
we follow Ref.~\cite{Mas:08}). Due to the strong interaction of the atoms
with the light field, the collective 
back action of the atoms on the light field is significant and the light
field dynamics becomes strongly dependent on the dynamics of the atoms. In
turn, the atomic dynamics is strongly 
driven by the light field and hence  an inherent strongly non-linear collective dynamical behavior arises. An approach by treating the interatomic 
interaction in terms of an effective static mean field \cite{Nag:08}, which is 
very successful for the pure BH model \cite{Lew:12}, yields useful first
insights. However, the 
collective interatomic interaction is an intrinsically time-dependent
non-equilibrium quantity and 
hence a more refined treatment is necessary. 
 
In this Letter, we use the bosonic dynamical mean field theory (BDMFT, see
\cite{Sno:13}) in which the quantum many-body
interaction dynamically evolves and 
eventually can be evaluated self-consistently. Due to its non-perturbative
nature, BDMFT is able to capture the strong dynamical back action induced
by the cavity, as has been 
demonstrated in the pioneering work in Ref.~\cite{Li:13}. An important scope
of the present paper is to adapt BDMFT to model the realistic experimental
scenario of Fig.~\ref{fig:Fig.1} and 
present calculations of experimentally accessible observables versus the
frequency and strength of the pump, which are the relevant tunable
parameters in experiments. Our 
description of the pumped optical cavity includes the unavoidable cavity
dissipation and the associated fluctuations due to the coupling to the
bath of vacuum radiation modes. Below a 
critical  pump we find a normal phase consisting of
the conventional optical lattice formed by the 
external periodic potential, which remains unperturbed by pump photons 
scattered into the cavity in an uncorrelated manner. 
Above the critical pump, we find a SO phase, related to what 
has been observed in experiments without an external lattice. 
This SO phase is superfluid, i.e., both the SO and the superfluid order
parameters are nonvanishing. At a second critical value of the pump 
strength, the system turns into a Mott insulator for even  
higher pump strengths. Our simulation results are compatible with a square root growth of the order 
parameter, (i.e. a critical exponent of 1/2) as function of the pump strength as also found in Refs. \cite{Nag:08, Nie:11}. We calculate the superfluid and SO order
parameters, the depths of the associated effective intra-cavity optical
potentials, the shape of the corresponding atomic 
density distributions, and study the dependence of the critical pump
strengths on the values of the cavity decay rate and the pump-cavity
detuning.


\textit{Generalized Bose-Hubbard model - }  We consider a BEC of $N$
identical atoms of mass $m$ trapped in an external periodic potential in
the $xz$-plane with the motion along the $y$-
direction assumed to be frozen out by a strong harmonic confinement with
the harmonic frequency $\Omega$ given by $\hbar \Omega \approx
10\,E_\textrm{rec}$ with $E_\textrm{rec}\equiv 
\hbar^2 k^2/2m$ denoting the recoil energy, and $k=2\pi/\lambda_p$. Hence,
a description in terms of a conventional two-dimensional BH model with
regard to the $x$ and $z$ coordinates is 
justified. This system is coupled to a standing wave cavity mode of
frequency $\omega_{c}$. The linear cavity is along the $x$-axis,
characterized by a field decay rate $\kappa$, and 
pumped by a transverse optical standing wave along the
$z$-direction with a strength $V_p$ and the wavelength $\lambda_{p}$.
Assuming the bad cavity limit $\hbar\,\kappa \gg 
E_\textrm{rec}$, such that $\kappa$ sets the fastest time scale
\cite{Gar:14}, we may follow Ref. 
\cite{Mas:08} and adiabatically eliminate the light field. By this, we
may replace the photon field operator by a complex valued coherent field
amplitude $\alpha$, which is a 
function of the matter variables, and write the effective Hamiltonian
\cite{Mas:08, Li:13}
\begin{eqnarray}\label{H}
\label{eq:eq_1}
\mathcal{H}&=&-\sum_{\langle i,j\rangle}
\tilde{J}_{ij}\,b^\dagger_ib_j+\frac{U}{2}\sum_i \hat{n}_i(\hat{n}_i-1) 
\\
&+&2\,\mathrm{Re}(\alpha)\,\eta_\textrm{eff}\,\tilde{J}_0\sum_i
(-1)^{i+1}\hat{n}_i+\sum_i (V_i^\mathrm{trap}-\mu_\mathrm{eff})\,
b^\dagger_ib_i \nonumber \, .
\end{eqnarray}
In passing, we note that the elimination of the light field is a very well-justified approximation
 for the setup of Ref.~\cite{Bau:10} but not for those of
Refs.~\cite{Kes:14, Wol:12}, where the photons remain much longer inside
the cavity before they are lost, such that an explicitly dynamical
treatment of the light field is required. The first line in
Eq.~(\ref{eq:eq_1}) exhibits a conventional BH Hamiltonian with
bosonic annihilation (creation) operators for the atoms $b_i
(b^\dagger_i)$, occupation number operators $\hat{n}_i=b^\dagger_i\,b_i$ at site
$i$ and the total particle number $N=\sum_i^{N_{s}} n_i$, where
$N_{s}=n_{L}^2$ is the total number of sites in a $n_{L}\times n_{L}$
square lattice with the linear lattice size $L=n_{L} \lambda_{p}/2$. The
nearest neighbor hopping amplitudes are $\tilde{J}_{ij}=\tilde{J}_x$ and
$\tilde{J}_{ij}=\tilde{J}_z$ for hopping in the $x$- and $z$-direction,
respectively, and $U$ denotes the on-site collision energy per particle. A
key element of $\mathcal{H}$ is the 3rd term with the effective 
pump laser strength $\eta_\mathrm{eff}=-\sqrt{|U_0| V_p}$, the light-shift
per photon $U_0$, and $\alpha=\eta_\textrm{eff}\tilde{J}_0
\sum_i(-1)^{i+1}\,n_i /(\Delta_c +i\kappa)$, where
$\Delta_c=\omega_p-\omega_c - U_0 J_0 N$ denotes the effective pump-cavity
detuning. The site-independent on-site matrix elements $J_0$ and
$\tilde{J}_0$ associated with scattering photons within the cavity and
between the pump and the cavity, respectively, follow after an expansion
in terms of Wannier states. In the 4th term the effective chemical
potential depends on the state of the light-field and reads
$\mu_{\textrm{eff}} = \mu_0-U_0 |\alpha|^2 J_0$. This directly expresses
the interdependence between the state of the light field $\alpha$ and the
chemical potential  (i.e., the energy scale for adjusting the particle number), which is
the origin of the nonlinear behavior of this system \cite{Mas:05}. The fourth
term also accounts for an additional external trapping potential shifting
the energy of the site $i$ by an amount $V_i^\mathrm{trap}$.

\begin{figure}
\begin{center}
\includegraphics[width=\linewidth]{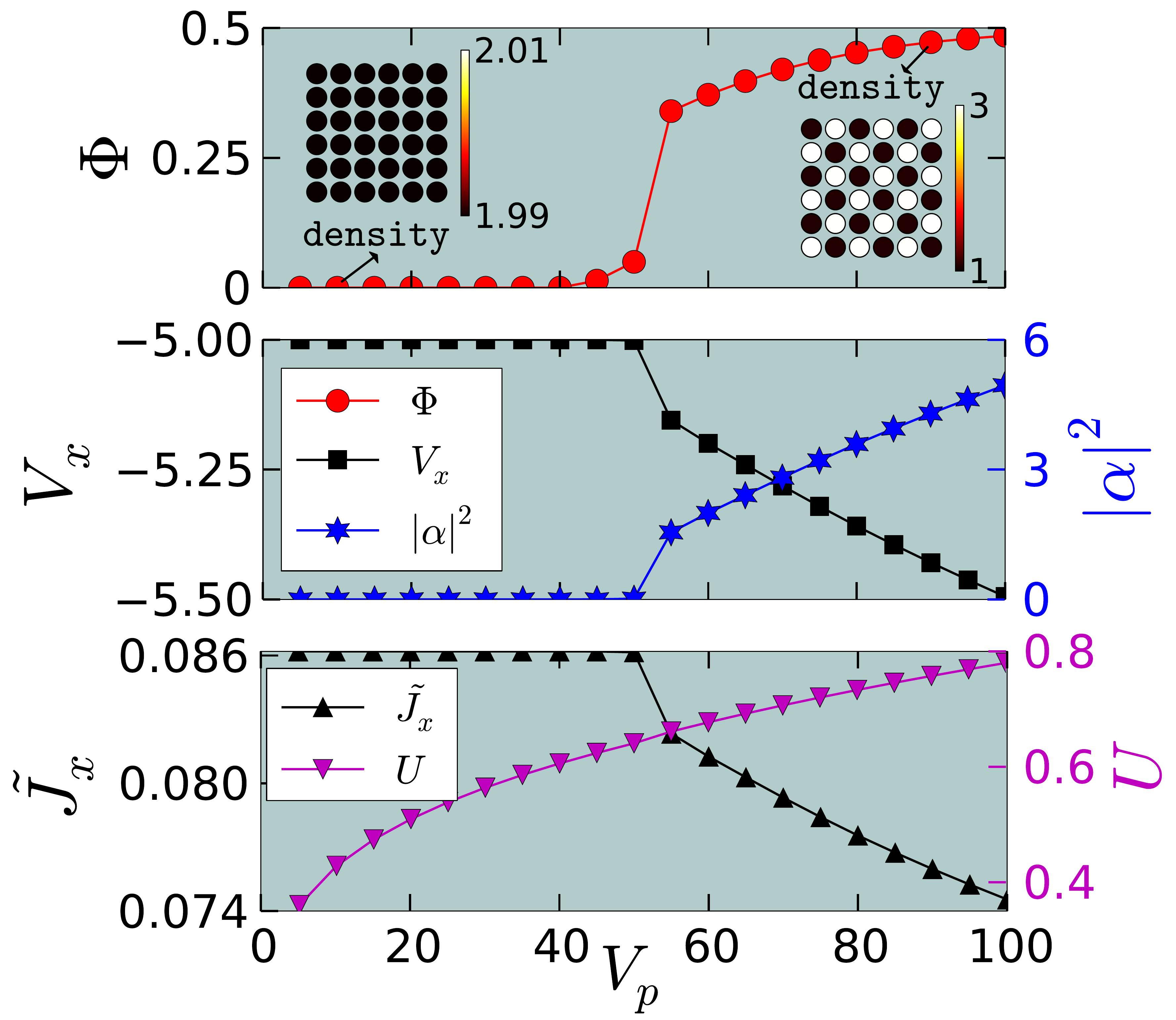} 
\caption{Top: Order parameter $\Phi$ of the self-organized phase and two
exemplary density distributions. Middle: Effective potential
$V_\mathrm{eff}\equiv V_x=V_\mathrm{cl}+U_0|\alpha|^2$ along the cavity axis (left
ordinate) and mean photon number $|\alpha|^2$ in the cavity (right
ordinate). Bottom: Resulting hopping amplitude $\tilde{J}_x$ along the
cavity ($x$) axis (left ordinate) and calculated on-site interaction $U$
(right ordinate). All panels are vs the transverse pump laser strength
$V_p$ (in units of $E_{\textrm{rec}}$). The cavity decay rate, detuning
and external classical potential are $\kappa=50\, E_{\textrm{rec}}$,
$\Delta_c=-2.8\, E_{\textrm{rec}}$ and
$V_\mathrm{cl}=-5\,E_{\textrm{rec}}$. }
\label{fig:Fig.2}
\end{center}
\end{figure}

Unlike for the standard BH model, the hopping amplitudes $\tilde{J}_{x,z}$
and the on-site interaction $U$ are not fixed \textit{a priori}, but are
given in terms of the instantaneous state of the photon field. In
particular, in harmonic approximation one gets
$\tilde{J}_{x,z}=\frac{4}{\sqrt{\pi}}(\frac{V_{x,z}}{E_{\textrm{rec}}})^{
3/4}\, \exp(-2\sqrt{V_{x,z}/E_{\textrm{rec}}}\,)\, E_{\textrm{rec}}$ and
$U=4\sqrt{2\pi}(\frac{a_s}{\lambda_p})\,(\frac{V_x V_z \hbar^2 \Omega^2}{4
E_{\textrm{rec}}^4})^{1/4}\,E_{\textrm{rec}}$ with $V_z=V_p$,
$V_x=V_{\textrm{cl}}+U_0\,|\alpha|^2$ and $V_{\textrm{cl}}$ denoting the
depth of the external lattice potential. The effective potential $V_x$
along the cavity axis explicitly depends on the photon number, showing
that the effective lattice potential inside the cavity is dynamically
formed. For a realistic description, we set
$V_{\textrm{ext}}=-5\,E_{\textrm{rec}} $ to ensure the validity of the
tight-binding picture. We set the  scattering length to $a_s=5.77
\,\mathrm{nm}$ and the pump laser wave length to 
$\lambda_p=803\, \mathrm{nm}$ according to  the setup in Refs.~\cite{Wol:12,
Kes:14}. In what follows, we focus on the homogeneous system and set the
external trap potential to zero, i.e., $V_i^\mathrm{trap}=0$. Moreover, we
consider a given light shift of $U_0=-0.1 \,E_{\textrm{rec}}$ at fixed
temperature of $k_B T=0.1\, E_{\textrm{rec}}$. We have checked that this value
of $T$ is low enough to exclude any finite-temperature effect as discussed, e.g., 
 in Ref.~\cite{Pia:13}.

 
\textit{Method -} To compute the resulting steady state of $\mathcal{H}$,
we use BDMFT (see also Ref.\ \cite{Li:13}) which is analogous to the fermionic
counterpart \cite{Vol:12}
and which is a nonperturbative approach to study a strongly correlated
many-body bosonic system.  The reliability of BDMFT depends on the
behavior of the BH model in the limit of $z\to\infty$ where $z=2d$ is the
coordination number in a $d$-dimensional lattice \cite{Sno:13}. In this
limit, one can rigorously show  that the BH model retains a local
many-body self-energy. As a net result, the BH model is mapped onto an
effective bosonic impurity model. Furthermore, a real-space extension of
BDMFT has been developed recently which importantly incorporates any
site-dependent behavior, either due to an external trapping potential
\cite{Li:11, Li:12}, or, solely by the underlying physics like the
existence of a self-organized phase as in the current case. We eventually
perform an exact diagonalization of the Anderson impurity model with
fairly small numbers of 
orbitals ($n_s=4-6$) to obtain the local densities $n_i$. Simultaneously,
we monitor also the cavity photon number $|\alpha|^2$ until convergence
for both parts is achieved. Throughout this work,  we consider a fixed total particle number $N\simeq 72$ on a $6\times6$
lattice. Even though our main physical finding is captured sufficiently well with this lattice size, however finite size effect leads to minor modification of phase boundaries e.g in Fig.~\ref{fig:fig4} . To quantify the self-organized quantum phase transition, we calculate the order parameter
as \cite{Nag:08, Li:13} $\Phi=\sum_i (-1)^i n_i / \sum_i n_i$. In a
perfect conventional optical lattice with site-independent $n_i$, we get
$\Phi=0$, while for a perfect checkerboard density pattern with $n_i = \pm
1$, we have $\Phi=1$. 

\textit{Results -} In the top panel of Fig.~\ref{fig:Fig.2}, we show the
order parameter $\Phi$ of the self-organized phase vs.\ the transverse
pump laser $V_p$. A clear quantum phase transition can be
observed. A critical pump laser strength $V_p^\mathrm{crit}$ exists, below
which  the system is a conventional optical lattice with $n_i\simeq 2$. In
other words, for $V_p<V_p^\mathrm{crit}$ the BEC atoms are
homogeneously distributed across the potential minima of the external
lattice, which exhibits square geometry with $\lambda_p/2$ separation of
adjacent sites along the $x$- and $z$-directions, as is illustrated in the
inset on the left-hand side of the upper panel. The scattering of pump
photons into the cavity from atoms at adjacent lattice sites interferes
destructively, such that the intra-cavity photon field $\alpha$ vanishes
and the effective lattice depth $V_x$ is thus given by the external
lattice depth $V_\mathrm{cl} =-5\,E_{\textrm{rec}}$, as is shown in the
middle panel of 
Fig.~\ref{fig:Fig.2}. As is seen in the lowermost panel, the resulting
hopping amplitude $\tilde{J}_x$ along the cavity axis is only determined
by the external lattice and hence constant, while $U$ grows monotonously
due to the growing confinement in the $z$-direction resulting from the
increasing intensity of the pump.

Upon increasing $V_p$, a pronounced quantum phase transition occurs. The
BEC atoms, previously homogeneously distributed across all sites of the
external lattice, self-organize to 
populate only every second site of the external lattice such that the
Bragg condition for coherent scattering of pump light into the cavity is
satisfied. Hence, a stationary intra-cavity 
field $|\alpha|$ emerges (see the middle panel in Fig.~\ref{fig:Fig.2}),
which acts to deepen the potential wells at every second lattice site and
thus stabilizes the self-organized 
density wave. The modulus of the associated effective potential $V_x$
along the cavity axis increases (middle panel) and hence the atom hopping
along the cavity axis $\tilde{J}_x$ is 
progressively impeded (lower panel). The order parameter $\Phi$
quantifying this self organization phenomenon (plotted in the top panel)
acquires non-zero values. Note that the density 
modulation is between $n_i=1$ and $n_i=3$, such that we obtain $\Phi<1$
even deeply within the SO phase. 
The corresponding real-space checkerboard-like density distribution is
shown in the inset on the right-hand side. Depending on initial thermal or
quantum fluctuations the SO density wave 
can arise with two spatial phases shifted with respect to each other by
$\lambda_p/2$ along the $x$-axis, i.e., the system faces a symmetry
breaking when crossing $V_p^\mathrm{crit}$. The 
SO phase transition also occurs for a vanishing depth of the external
lattice, in which case it represents the well-known Dicke phase transition
\cite{Gro:82, Aga:12} that has been observed 
in recent experiments \cite{Bau:10, Kes:14}. A related result for $\Phi$
vs $V_p$ was obtained by a static mean-field approach \cite{Nag:08}, but
we cannot quantitatively compare it with 
our results since in BDMFT the interactions are not fixed \textit{a
priori}, but are calculated dynamically and self-consistently. We note that in
the regime of strong pumping, regions in the phase diagram have been identified
within a strong coupling expansion where the phases show a finite
compressibility \cite{Vid:10}. 

\begin{figure}[t!]
\centering
\includegraphics[width=\linewidth]{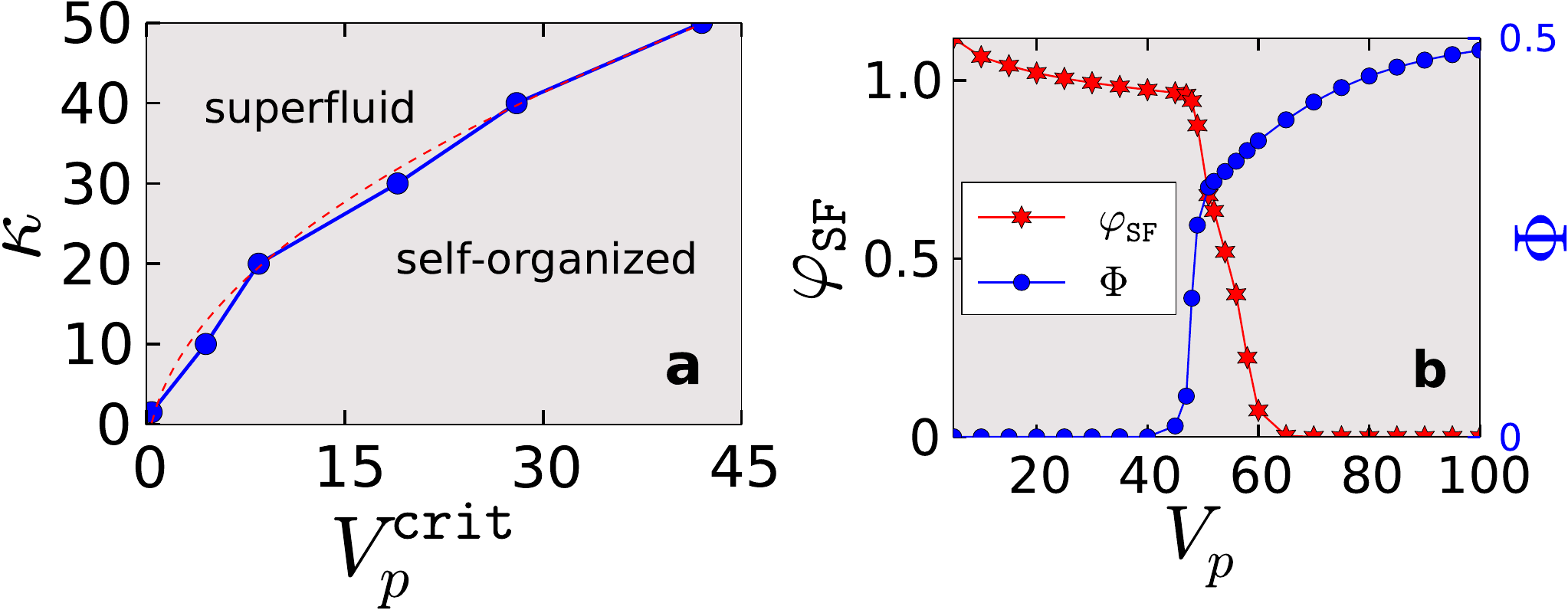} 
\caption{(a) Phase boundary between the superfluid and the self-organized
phase for varying cavity decay rates $\kappa$ and critical pump laser
strengths $V_p^\mathrm{crit}$ (both in 
units of $E_{\textrm{rec}}$). The dashed line represents a square-root fit.
(b) Superfluid order parameter $\varphi_\mathrm{SF}$ (left ordinate) and
SO order parameter $\Phi$ (right ordinate) plotted against the pump
strength $V_p$ (in units of 
$E_{\textrm{rec}}$). The cavity decay rate and the detuning are set to
$\kappa=50\,E_{\textrm{rec}}$ and  $\Delta_c=-2.8\,E_{\textrm{rec}}$,
respectively.}
\label{fig:Fig.3}
\end{figure}

In Fig.~\ref{fig:Fig.3}(a), we show the dependence of
$V_p^\mathrm{crit}$ on $\kappa$ and compare our calculations with a
square-root fit as qualitatively suggested in Ref.\ \cite{Kul:13}. Note
that an increase of $\kappa$ while $U_0$, and, hence the cavity finesse, is
kept constant amounts to a decrease of the cavity length. The real-space
BDMFT allows us also to calculate the superfluid order parameter
$\varphi_\mathrm{SF}=\langle b\rangle$ \cite{Li:13}. This is depicted in
Fig.~\ref{fig:Fig.3}(b) together with $\Phi$. When $\Phi$ departs from
zero at $V_p^\mathrm{crit}$, $\varphi_\mathrm{SF}$ begins to decrease.
Interestingly, there is a sizable window in the vicinity of
$V_p^\mathrm{crit}$ where $\varphi_\mathrm{SF}$ and $\Phi$ simultaneously
coexist with a nonzero value, which is a manifestation of long-range superfluid
order together
with an SO density-wave. One may be tempted to call this state a
supersolid phase \cite{Bon:12, Li:13}. However, it should be noted that translational symmetry of the effective lattice, which traps the SO 
pattern, is not broken here because the lattice has an effective periodicity of $\lambda$ corresponding to the density modulation of the atomic state. 
Above a certain value of $V_p$, $\varphi_\mathrm{SF}$ vanishes, and the Mott insulating phase arises.

In Fig.~\ref{fig:fig4}, we depict the phase diagram of our system with
respect to $V_p$ and negative pump-cavity detuning $\Delta_c < 0$ for a
fixed cavity decay rate $\kappa=50\,E_{\textrm{rec}}$. The phase boundary
between the normal phase and the self-organized superfluid phase (SO/SF)
is found in accordance with the experiments in Refs.~\cite{Bau:10,
Kli:14}. For small $|\Delta_c|<100 \,E_{\textrm{rec}}$ (see inset), the
critical pump power scales inversely with $|\Delta_c|$ while for large
values of $|\Delta_c|$, it scales linearly with $| \Delta_c |$. The notable
increase of $V_p^\mathrm{crit}$ as $|\Delta_c|$ falls below $\kappa$ is
explained as follows: the intra-cavity field is driven by the pump field
with a phase delay approaching $\pi/2$ as $\Delta_c$ approaches zero and
hence the two fields cease to interfere. Hence, the resulting light field
acquires potential minima of equal depth at each potential minimum of the
external potential. The checkerboard density wave pattern is then no
longer supported 
and hence the scattering of light into the cavity is suppressed. Note that
a similar phase boundary is found for large atom samples with no external
lattice described by a semi-classical kinetic Vlasov equation which
neglects all correlations \cite{Nie:11}. Even though our particle number
is much smaller than in Ref.~\cite{Nie:11}, we still correctly
capture the qualitative mean field physics. A second phase boundary occurs,
where the self-organized phase becomes a Mott insulator (SO/MI).

\begin{figure}
\centering
\includegraphics[scale=0.35]{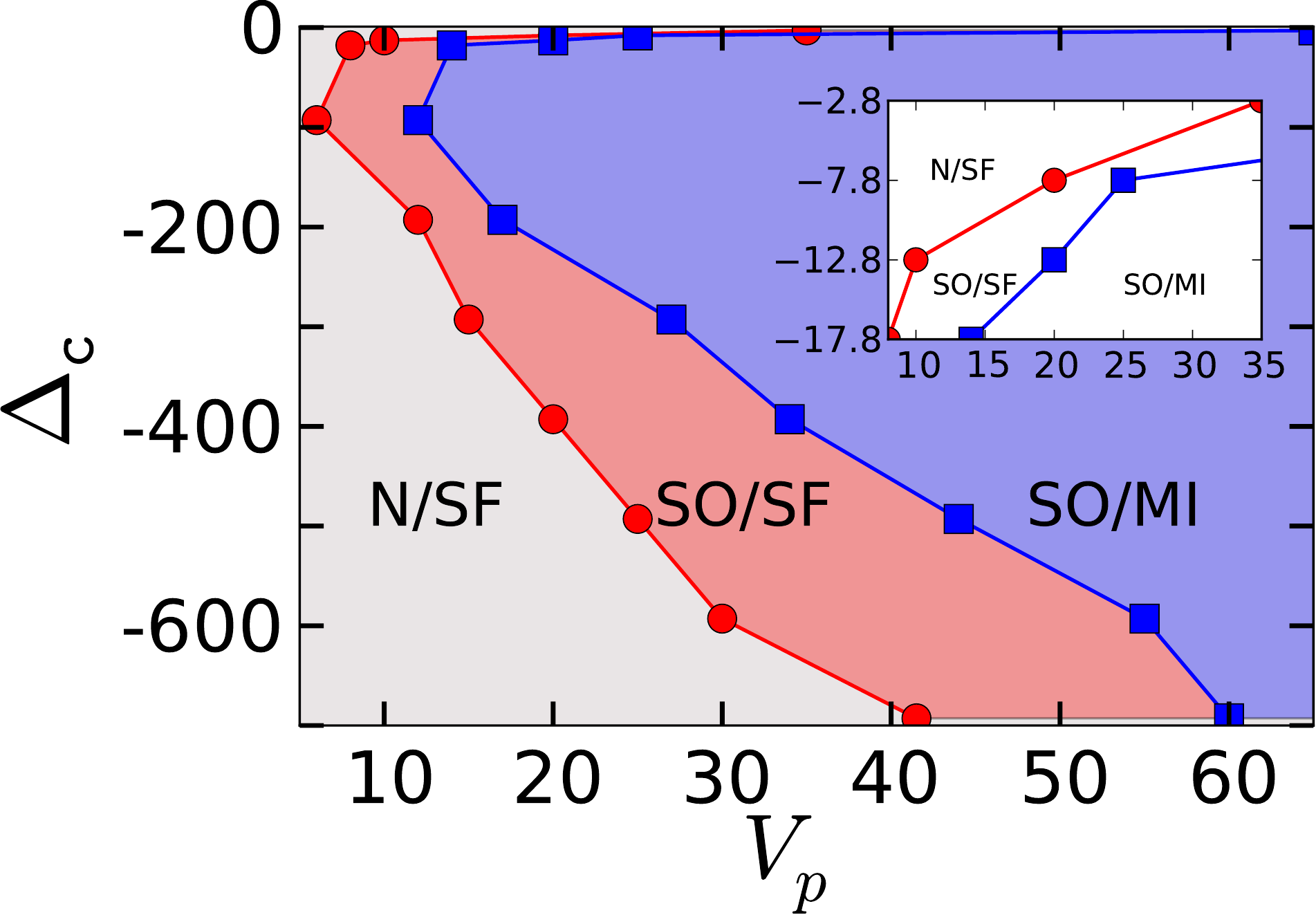} 
\caption{Phase diagram with respect to $\Delta_c$ and $V_p$ (both in units
of $E_{\textrm{rec}})$. N/SF refers to the normal superfluid phase, SO/SF
to the self-organized superfluid phase and SO/MI to the self-organized
Mott insulator. The inset shows a zoom of the main plot for small values
of $|\Delta_c|$ ($\kappa=50\,E_{\textrm{rec}}$).}
\label{fig:fig4}
\end{figure}

In conclusion, we have solved the generalized Bose-Hubbard model for an
optical lattice of bosons interacting collectively with an optical cavity
mode in a scenario adapted to recent experiments. Using the non-perturbative
real-space BDMFT, we find two nonequilibrium quantum phase transitions upon
increasing the strength of a transverse pump field. A quantum phase transition
from the normal phase to a self-organized superfluid phase occurs. For even
stronger pumping, a pure self-organized Mott insulator phase 
arises.  

\begin{acknowledgments}
This work was supported by the SFB925 (projects C5 and C8) and by the Austrian Science Fund project S4013. We acknowledge useful discussions with M.~Wolke, H.~Kessler, J.~Klinder,
J.~Larson, P.~Nalbach, L.~He and
Y.~Li. In particular, M.R.B. thanks Walter Hofstetter for his previous
major input on the familiarization with DMFT.  
\end{acknowledgments}

\end{document}